\documentstyle[aps,epsfig,twocolumn]{revtex}

\begin{document}

\title{Phase ordering and shape deformation of two-phase membranes}

\author{Y. Jiang~\cite{author1}, T. Lookman$^*$, and A.~Saxena }  
\address{Theoretical Division, Los Alamos National Laboratory, Los
  Alamos, New Mexico 87545 }
\address{$*$Department of Applied Mathematics, University of Western
Ontario, London, Ontario, N6A 5B7 Canada}

\maketitle

\begin{abstract}
Within a coupled-field Ginzburg-Landau model we study analytically
phase separation and accompanying shape deformation on a two-phase
elastic membrane in simple geometries such as cylinders, spheres and
tori.  Using an exact periodic domain wall solution we solve for the
shape and phase ordering field, and estimate the degree of deformation
of the membrane.  The results are pertinent to a preferential phase
separation in regions of differing curvature on a variety of vesicles. 
\end{abstract} 

\pacs{Pacs numbers: 87.22.Bt, 64.60.Cn, 11.10Lm} 

Amphiphilic molecules assemble in aqueous media to form bilayers,
which close to form vesicles at low concentrations.  Bilayers and
vesicles serve as models for membranes and cells for studying simple
physical properties such as shape deformations, elasticity and
transport.  They show an amazing variety of shapes, which have been
described by treating the membrane as a homogeneous elastic sheet
with area and volume constraints~\cite{canham70,helfrich}.
However, recent experimental observations have recognized that
internal degrees of freedom can crucially influence the shapes.  An
example is the transition from a normal biconcave shape of {\em
  discocytes\/} to a crenated shape of {\em echinocytes\/} of a 
human red blood cell.  Such transformations can be induced by an 
asymmetric adsorption of certain drugs, i.e., a  local asymmetry in
the composition plays an important role in this crenated
shape~\cite{sheetz}.  

As molecules are free to move in the plane of the membrane, lateral
phase separation is constantly observed in lipid membranes.  A single
component membrane under certain conditions can exhibit regions rich
in tilted and non-tilted phases, respectively, while a two-component
membrane can exhibit phase separation of both different components and
tilted vs. non-tilted phases \cite{sackmann77}.  Phase separation
plays a central role in the stabilization of vesicles and in the
fission of small vesicles after budding~\cite{lipowsky93,budding}.  
Although experimental studies that clearly relate phase separation
to local shape deformation \cite{sackmann77} are scarce, a number of 
phenomenological and numerical investigations have shown that a
coupling of the local curvature to the local composition of
amphiphiles can result in shape deformation 
~\cite{sackmann95,taniguchi9394,taniguchi96} and budding
~\cite{rao98}. 

The numerical studies have considered mainly a general fluctuating
vesicle and therefore the central role of the coupling between the
phase separation and accompanying shape deformation process has been
difficult to decipher.  Our work is thus motivated by a desire to
study, by analytical means where possible, phase separation on the
simplest of geometries such as cylinders, spheres and tori.  While
microtubules are abundant in biology, deformable spherical and
toroidal vesicles have also been observed~\cite{bensimon95}. We seek 
to provide a systematic formal description of the equilibrium
solution.  Our aim is to gain insight into the part played by
curvature and to extract the salient ingredients that affect phase
separation and shape changes.   We estimate the
degree of deformation from the coupling strength between the
composition and curvature fields and the elastic rigidity.  

We represent a membrane as a surface embedded in three dimensions
parameterized by ${q}\equiv\{q_1, q_2\}$, for its thickness is usually
several orders of magnitude smaller than its size.  Our approach is to
study phase separation on a subset of surfaces (orthogonal curvilinear
manifolds \cite{arf}) which have either an axis of translation or
rotation.  For this special class of surfaces we have recently derived
some simplifying analytical results regarding phase separation on {\it
  rigid} curved surfaces~\cite{saxena99}.  Here we 
apply the analysis to deformable surfaces.  

The total free energy of the membrane is $F=F_1+F_2+F_3$, with the 
bending elastic energy~\cite{helfrich}:
\begin{equation}
F_1 = \int dA~\left[{\frac {\kappa}{2}} (h-h_0)^2 \right] ,
\end{equation}
\noindent where $dA=\sqrt{g}\/d^2q$ is the area element with
$\sqrt{g}$ the determinant of the metric tensor $g_{ij}$, $h$ the mean
curvature, and $h_0$ the spontaneous mean curvature--the preferred
curvature of the relaxed vesicle.  The mean curvature is
$h=h(q_1,q_2)= h_1+h_2$, where $h_1$ and $h_2$ are the principal
curvatures.  For an arbitrary surface embedded in three dimensions,
if $q_1$ and $q_2$ are orthogonal coordinates, the metric tensor has
$g_{ij}=0$ for $i\neq j$ and $\sqrt{g}=\sqrt{g_{11}g_{22}}$. 

We treat phase separation within the usual Ginzburg-Landau free energy 
framework:
\begin{equation}
F_2 = \int dA~\left[ {\frac{\xi^2}{2}}| \nabla \phi |^2 + V(\phi)\right ]  ,
\end{equation}
\noindent where $\phi$ is either the relative concentration of the two
phase components $A$ and $B$ of the membrane: $\phi =(A-B)/(A+B)$, or
the concentration of a diffusing external chemical, as in the case of 
{\em echinocytosis } of red blood cells \cite{sheetz,leibler86}.    
Here $\xi$ is the characteristic length, which determines interface
width.   

We use the bilinear form of coupling between the phase density and
local curvature~\cite{leibler86}, an interaction energy found on
phenomenological grounds: 
\begin{equation} 
F_3 = \int dA~\Lambda \phi h ,
\end{equation}
\noindent where $\Lambda$ is the strength of the coupling.

More realistic considerations should take into account area and volume
constraints for vesicle membranes.  A change in area would increase the
surface energy and a change in volume would increase the osmotic
pressure.  Hence an additional term
\begin{equation}
F_4 = \lambda A +PV= \int dA~(\lambda +P{\frac {|r|}{3}}) , \nonumber 
\end{equation} 
\noindent where $\lambda$ is the surface tension and $P$ the osmotic
pressure, should be included in the free energy $F$.  The second term  
in the expression utilizes the divergence theorem in 3D.  The surface
tension $\lambda$ is a constant and does not enter the variational 
calculations.  The relation between the local radius $|r|$ and the
mean curvature $h$ can be highly complicated depending on the
geometry, which will render the free energy too intractable for our
purposes.  Therefore, we consider only $P=0$ hereafter.

In the case where the order parameter is conserved, as we consider no
exchange of molecules of the membrane with its environment, an
appropriate equation for the time evolution of the phase ordering field 
$\phi$ is:
$${\frac {\partial \phi } {\partial t}}  =  \Gamma_1 \nabla_{LB}^2
{\frac {\delta F}{\delta \phi}}  \nonumber \\    
 =  \Gamma_1 \nabla_{LB}^2 \left [ V^\prime(\phi)-\nabla_{LB}^2\phi +
\Lambda h \right ] , $$ 
where prime denotes differentiation with respect to $\phi$.
Similarly, for the mean curvature: 
$${\frac {\partial h }{\partial t}}  =  \Gamma_2 \nabla_{LB}^2 { \frac
 {\delta F}{\delta h}}  \nonumber 
\\  =  \Gamma_2 \nabla_{LB}^2 \left [ (\kappa(h-h_0)+\Lambda\phi \right],
$$ 
\noindent where $\Gamma_1$ and $\Gamma_2$ are the diffusion
coefficients for $\phi$ and $h$ fields, respectively, and
$\nabla_{LB}^2 \equiv {\frac {1}{\sqrt{g}}}{\frac {\partial}{\partial
x^i}} (g^{ij}\sqrt{g}{\frac {\partial}{\partial x^j}})$ is the
Laplace-Beltrami operator.  As the details of the double-well
potential are not important~\cite{bray94}, we use a simple $\phi^{4}$
potential to describe the phase separation dynamics:
\( V(\phi)= {\frac {\alpha}{4}} \phi^4 - {\frac {\beta}{2}}
\phi^2 ,~ (\alpha, \beta > 0 ) . \)

The numerical study of the dynamics of evolution will be reported
elsewhere~\cite{jiangfuture}.  Here we focus on the 
analytic equilibrium solutions.  At equilibrium,
the Euler-Lagrange (EL) equations for $\phi$ and $h$ fields are 
${\delta F}/{\delta \phi} = 0 $ and $ {\delta F} /{\delta h} =0$,
respectively.  The EL equations are nonlinear and usually do not have
an exact closed-form solution.  In order to obtain analytical results,
we consider special symmetries to reduce the problem to a quasi-one
dimensional one.  If $q_1$ is the axis of symmetry, $|\nabla \phi| =
|{d\phi}/{dq_2}|$.  We then define a new variable $\tau_1$ as 
$d\tau_1 \equiv \sqrt{g_{22}/g_{11}}~dq_2$.
\noindent With this new variable, $\nabla_{LB}^2 \phi = 
{\frac {1}{g_{11}}}{\frac {d^2}{d{\tau_1}^2} \phi}$.  

The equilibrium condition, derived from the EL equation for $h$ is 
\begin{equation}
h = h_0 - {\Lambda \over \kappa} \phi,
\label{hphi}
\end{equation}
\noindent i.e., at equilibrium the local mean curvature of the membrane
is linearly proportional to the local $\phi$.   This linear
relationship explains why phase separated regions have local 
curvature $(\Lambda/\kappa)\phi$, a result that appeared in the 
numerical study of Ref.~\cite{taniguchi96}.  Thus, we can eliminate  
$h$ from the free energy.  It follows that the EL equation for the
free energy with respect to $\phi$ becomes
\begin{equation}
V_e^\prime(\phi) - {d^2 \over {d\tau_1^2} }\phi = 0  .
\label{elphi}
\end{equation}
\noindent Here $V_e$ is the new effective potential:
\begin{equation}
V_e \equiv g_{11}\left[ {\frac {\alpha}{4}}\phi^4 - {\frac {1}{2}}
  (\beta+ {\Lambda^2 \over \kappa}) \phi^2 + \Lambda h_0\phi \right] ,
\label{ve}
\end{equation}
\noindent which depends only on $\phi$.  The coefficient of
the  $\phi^2$ term is renormalized and the effective potential becomes
an asymmetric double-well due to the linear coupling.  

Twice integrating Eq.~(\ref{elphi}) we obtain a general periodic 
domain-wall lattice solution:
\begin{equation}
\phi(\tau) = d + {{c-d} \over {1- {{b-c} \over {b-d}} 
{\text {sn}}^2\left({{\tau-\tau_0} \over {\zeta}}, k \right)}} , k \equiv
\sqrt {\frac {(b-c)(a-d)}{(a-c)(b-d)}} ,
\label{phi}
\end{equation}
\noindent where $d\tau \equiv d\tau_1 \sqrt{g_{11}}=\sqrt{g_{22}}dq_2$
is the arc length variable; $a, b, c$ and $d$ are real roots of $V_e -
V_0 =0$, i.e. \(V_e - V_0 = g_{11}(\phi-a)(\phi-b)(\phi-c)(\phi-d) \)
with $a > b\geq \phi > c > d$; $V_0$, $\tau_0$ are two constants of
integration, $\zeta \equiv \xi \sqrt{4/\alpha}{\sqrt{2 /{(a-c)(b-d)}}}$
is the rescaled characteristic length scale, $k$ is the modulus of the
Jacobian elliptic function ${\text {sn}}(\tau,k)$.  The shape of the
periodic solution depends on the modulus $k$, which in turn depends on  
all the parameters of the model and the boundary condition $V_0$.  The
value of $k$ ranges between 0 and 1.  For $k=0$, the Jacobian elliptic
function reduces to a sinusoidal function.  For $k=1$, ${\text
{sn}}(\tau,k)$ changes to a kink solution, $\tanh(\tau)$, which is no
longer periodic and thus is only allowed on an open geometry.  For a
closed geometry, e.g. a circle, a periodic solution is required and
the number of periods for a fixed perimeter $L$ depends on the value
of $k$ \cite{snk}: the periods allowed should satisfy $L/\zeta=4mK$
where $ K(k)$ is the complete elliptic integral of the 
first kind and $m$ is an arbitrary integer.  With the linear
relationship between $\phi$ and $h$, Eq. (\ref{hphi}), we also 
obtain the expression for the mean curvature $h(\tau) = h_0 - {\frac
{\Lambda}{\kappa}} \phi(\tau)$.  

Transforming $\tau$ to space coordinate $q_2$ and exploiting the
fact that $q_1$ is the axis of symmetry, we obtain $\phi(q_1,q_2)$ and
$h(q_1, q_2)$.  From $h(q_1, q_2)$, we can then use the relationship
between $h$ and $r$, to obtain $r(q_1, q_2)$, the shape in real
space coordinates. 

For illustration purposes, we now carry out this formulation on a
cylinder with rotational symmetry $r=r(z)$.  The metric tensor has
$g_{\theta\theta}=r^2 $, $g_{zz}=1+{r^\prime}^2$ and $g_{\theta
  z}=0$. 

With $\theta$ as the axis of symmetry, we define the new
variable $\tau_1$ as
$$d\tau_1 = dz\/\sqrt{ {\frac {g_{zz}}{g_{\theta\theta}}} } =
 dz\/{\sqrt {\frac {r^2+{r^\prime}^2} {r^2} }} .
$$ 
\noindent  The free energy for such a cylinder is then 
\begin{equation}
F= \int d\tau_1 d\theta \/  \left [{\frac{\xi^2}{2}}\phi^2_{\tau_1} +
V_e(\phi)\right ] ,
\end{equation}
\noindent in which $V_e(\phi)$ is the same as Eq.~(\ref{ve}).  Applying
the EL equation for $h$, we obtain the same linear relation between
$h$ and $\phi$ as in Eq.~(\ref{hphi}) in terms of $\tau_1$.  Replacing  
$h$ in the free energy, $F$ becomes a function of $\phi$ only and its 
EL equation with respect to $\phi$ is
\[ {\frac{\delta F}{\delta\phi(\tau_1)}} = \xi^2 \phi_{\tau_1\tau_1} -
r^2 \left [\alpha \phi^3-(\beta + {\frac{\Lambda^2}{\kappa}})\phi
  +\Lambda h_0 \right ] = 0 . \]
\noindent This equation yields the same solution for $\phi$ as
Eq.~(\ref{phi}) as a function of the arc variable $\tau$: 
\begin{equation}
d\tau \equiv r~d\tau_1 = dz \/{\sqrt{1+{r^\prime}^2}} ,
\end{equation} 
\noindent  and hence the solution of $h(\tau)$ based on the linear
relationship between $\phi$ and $h$.

In order to convert the $\phi$ and $h$ results to phase distribution 
and deformation on a cylinder, i.e., to obtain $\phi(z,\theta)$ and
$r(z,\theta)$, we need to change the variable $\tau$ back to the
original variable $z$ as follows.  We replace $r=r(z)$ by $\rho(\tau)$,
such that  
$$\rho_{\tau} = r_z / \sqrt{1+{r_z}^2} , ~~~  
\rho_{\tau\tau} = r_{zz}/ (1+{r_z})^2 .
$$ 
\noindent  The mean curvature $h(z)$ in terms of $\rho(\tau)$ is
\begin{equation}
h(\tau)= 1/\rho + \rho_{\tau\tau}/\sqrt{1-{\rho_{\tau}}^2}.
\nonumber
\end{equation}
\noindent A numerical integration of this equation provides 
$\rho(\tau)$ and thus $z=\int d\tau\/\sqrt{1-{\rho_{\tau}}^2}$, which
in turn gives $r(z)=\rho(\tau(z))$, the deformation along the $z$ axis
of a cylinder.  

Figure 1a shows a typical plot for the equilibrium $\phi$, $h$ and $r$
as a function of $z$, obtained with the following parameters:
$\alpha=4$, $\beta=2$, $h_0=5$, $\xi=0.2$, $\kappa=0.02$, and
$\Lambda=0.03$.  Figure 1b is the corresponding {\it axially}  
deformed cylinder, whose shades of gray correspond to the amplitude of the 
order-parameter field.  Applying the same formulation to a cylinder
with an axial symmetry $r=r(\theta)$, we obtain a cylinder with
deformation occurring only along the cross section.  Figure 2a shows a
series of deformed circles with periods 3 to 6.  A {\it radially}
deformed cylinder, as shown in  Fig. 2b, is formed by translating the
deformed circle along the $z$ axis.  Figure 2b uses the same
parameters as in Fig. 1.   

The degree of deformation, defined as the ratio of the maximum
to the minimum radii, $D=r_{max}/r_{min}$, is a quantity that can be
measured by reflection interference contrast
microscopy~\cite{sackmann94} or atomic force microscopy~\cite{ermi98}.
We estimate the degree of deformation of an axially symmetric
cylinder, for which $r=1/H$ and thus
$D=({H_0-{\frac{\Lambda}{\kappa}}c})/({H_0-{\frac{\Lambda}{\kappa}}d})$,
as a function of $\kappa$ and $\Lambda$.   The value of $\kappa$ can be
either obtained by using micropipette 
methods~\cite{evans97}, or estimated from molecular dynamics
simulations~\cite{goetz99}.  Although difficult to measure directly in
experiments, $\Lambda$ may be derived from first principle molecular
dynamics simulations as in~\cite{goetz99} by using Eq.~(\ref{hphi}).
Figure 3a shows that the deformation $D$ has a quadratic form in
$1/\kappa$.  Figure 3b indicates again a quadratic form dependence of 
the deformation on the coupling constant $\Lambda$.  This can be
understood by expanding $D$ in the small deformation limit, when
$\Lambda \ll \kappa$.  In this limit, $c-d$ and $cd$ remain
approximately independent of $\Lambda$ and $\kappa$, thus $D \propto
c_1 + c_2\Lambda/\kappa+ c_3 \Lambda^2/\kappa^2$ with $c_1, c_2$ and $c_3$
constants.  Indeed, if we vary both $\Lambda$ and $\kappa$ such that
$\Lambda/\kappa$ remains constant, the deformed circles are virtually
identical (figure not shown).  

We also applied this framework to other deformable geometries and symmetries,
e.g. a sphere and a torus with appropriate $g_{ij}$.
Figures 4a,b and 4c,d show results for spheres and tori with axial and
azimuthal symmetry, respectively.  The parameters used are the same as
in Fig. 1.  We observe preferential phase separation similar to the
case of cylinders.

To summarize, we have developed a theoretical framework in which we
obtained exact analytical solutions for the equilibrium phase
distributions and membrane shapes, for a series of geometries
including cylinders, spheres and tori.  This framework allows for an
estimate of the degree of deformation from the coupling strength
and the elastic rigidity of the membrane.  Since fluid properties are
essential for modeling cells and membranes, our model augmented with a
coupling to hydrodynamics will enable the study of realistic
biological cells. 

The order-parameter field considered above need not be a scalar
density or relative concentration field.  If we choose magnetization
($M$) or polarization ($P$) instead of $\phi$, we would then expect
periodic stripes of magnetic and polarization domains in regions of
different curvature.    

We thank R.C. Desai and G. L. Eyink for useful discussions.  This
research is supported in part by the US Department of Energy, under 
contract W-7405-ENG-36.

\begin{figure}
\caption{Phase separation and deformation on a radially symmetric 
circular cylinder. (a) Plots of order parameter $\phi$, curvature 
$h$ and radius $r$ as a function of $z$.  (b) Equilibrium shape of the 
deformed cylinder; shades of gray correspond to the order-parameter 
$\phi$.}   
\centerline{ 
\epsfig{file=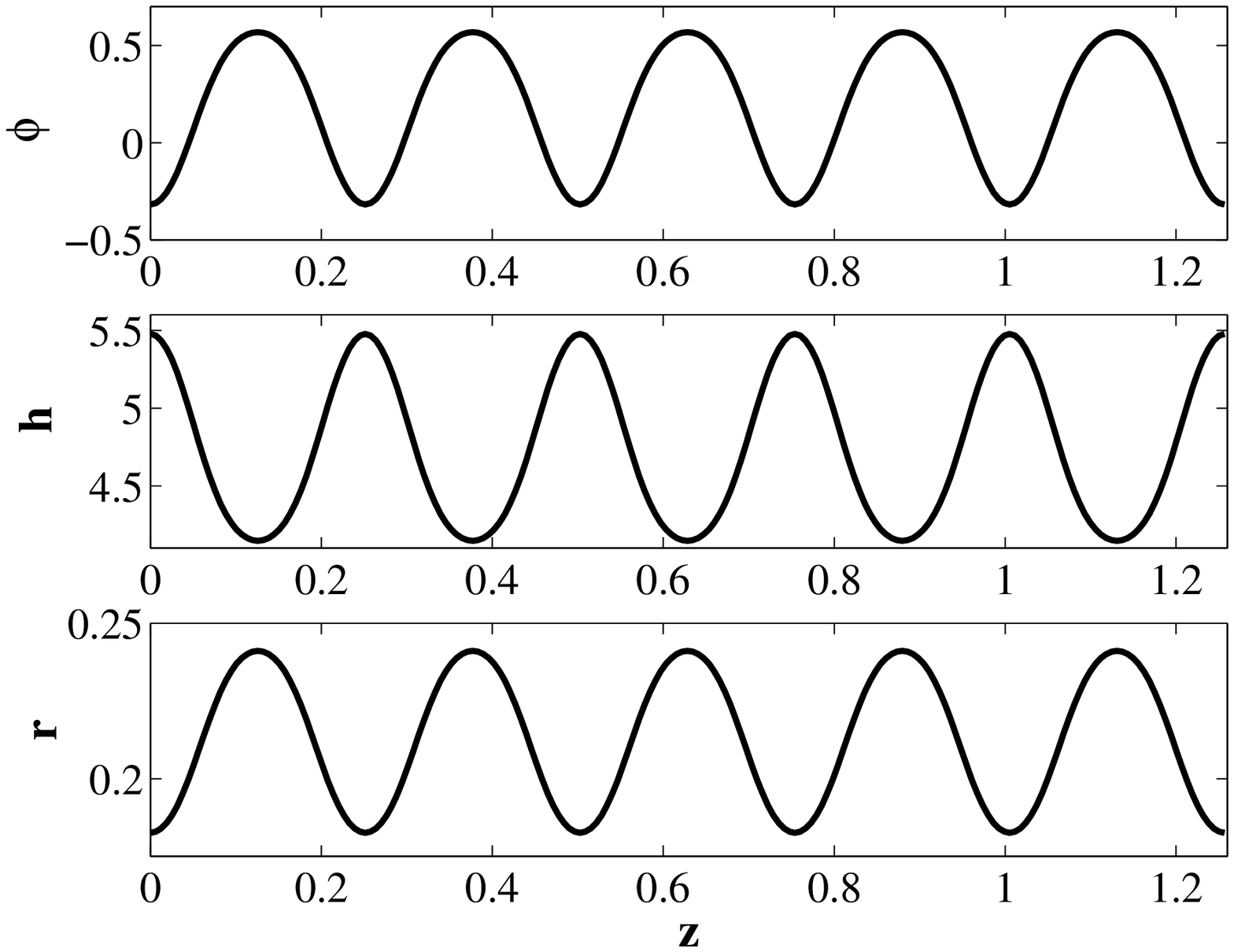, width=2.5in}
\epsfig{file=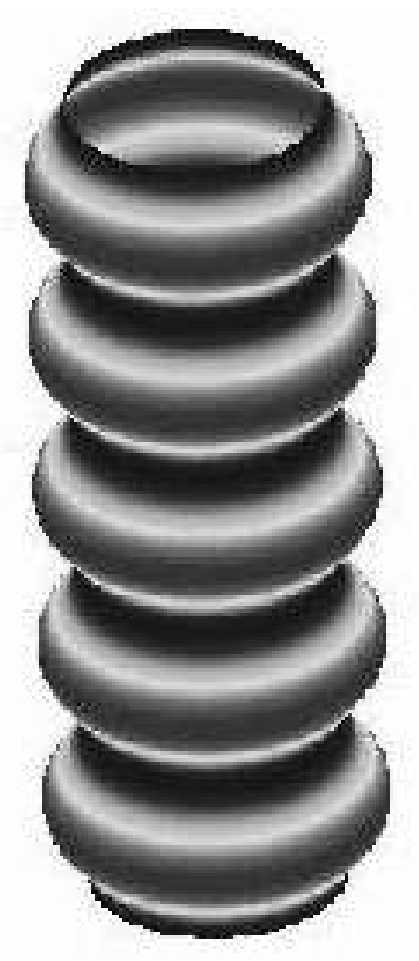, width=0.8in}}    
\end{figure}

\begin{figure}
\caption{Phase separation and deformation on an axially symmetric 
circular cylinder. (a) Cross sections of deformed cylinder with 3, 4, 5 
and 6 modes, respectively.  (b) Equilibrium shape of a deformed cylinder 
with mode 6; shades of gray correspond to the order parameter $\phi$.} 
\centerline{
\epsfig{file=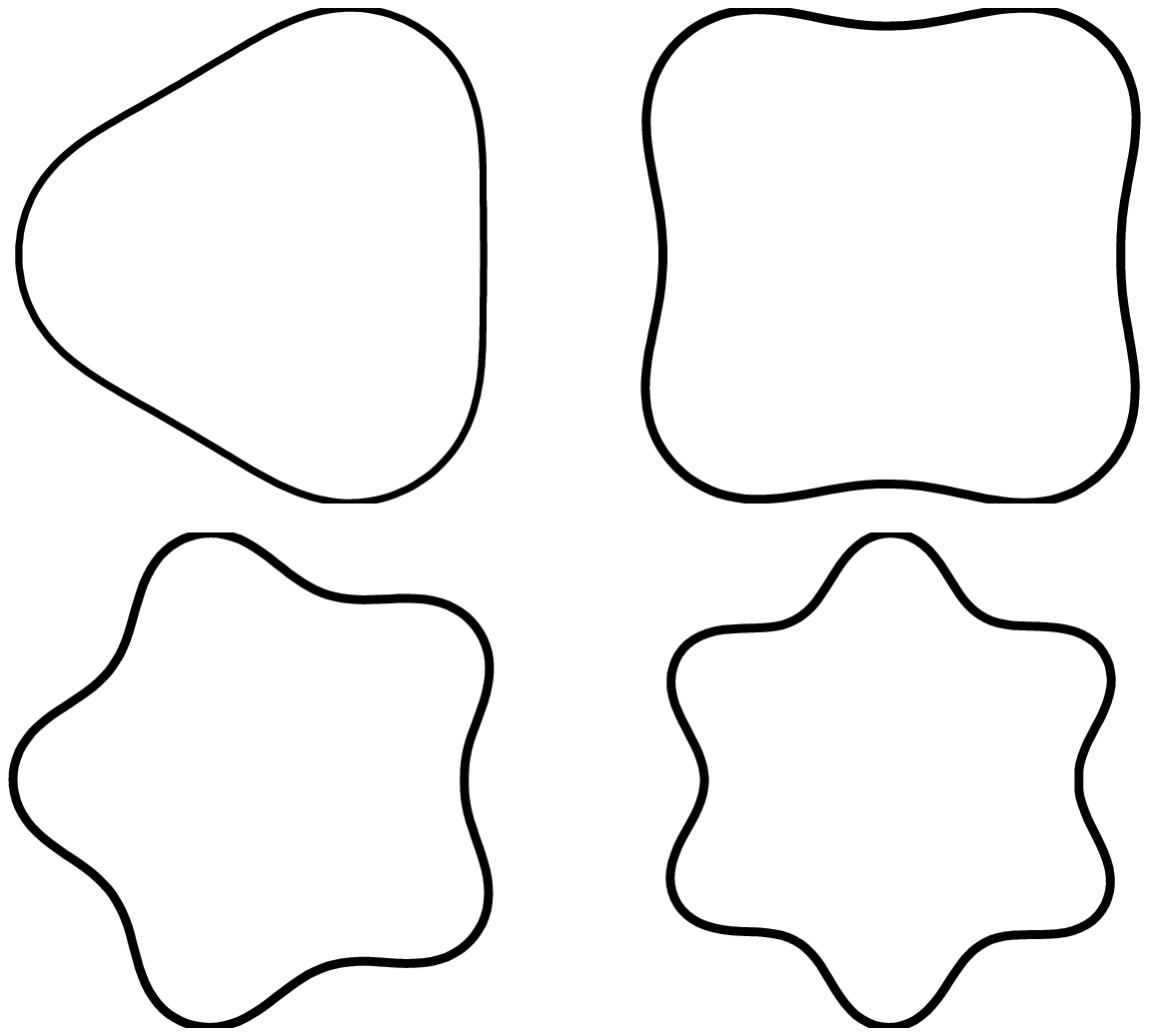, width=2.1in} 
\epsfig{file=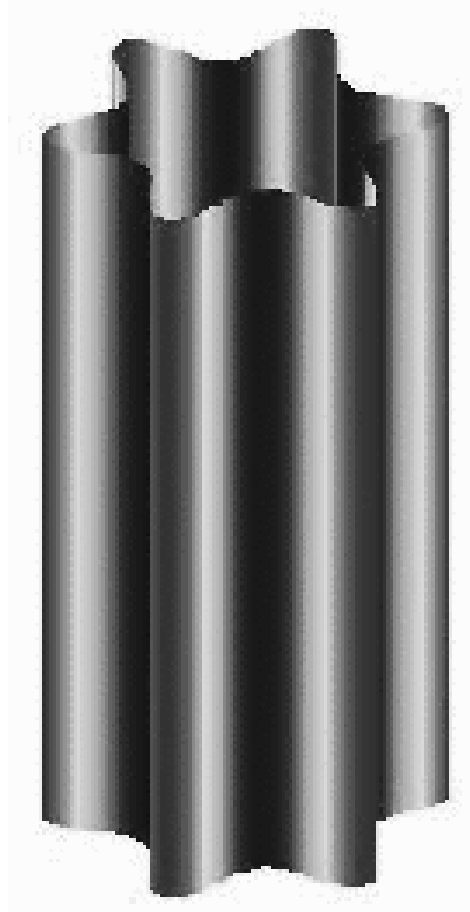, width=0.80in} }
\end{figure}

\begin{figure}
\caption{Deformation of the axially symmetric cylinder as a function
  of: (a) Elastic rigidity $\kappa$ for fixed $\Lambda=0.02$; 
circles are numerical data and the solid line is fit 
Inset shows the deformed cross sections of the cylinder, the
inner-most curve corresponding to smallest $\kappa$ value. (b)
Coupling constant $\Lambda$ for fixed $\kappa=0.01$; circles are
numerical data and the solid line is a fit to a quadratic form.  Inset
shows the deformed cross sections with fixed perimeter, the inner-most
curve corresponds to the largest $\Lambda$ value. }  
\centerline{
\epsfig{file=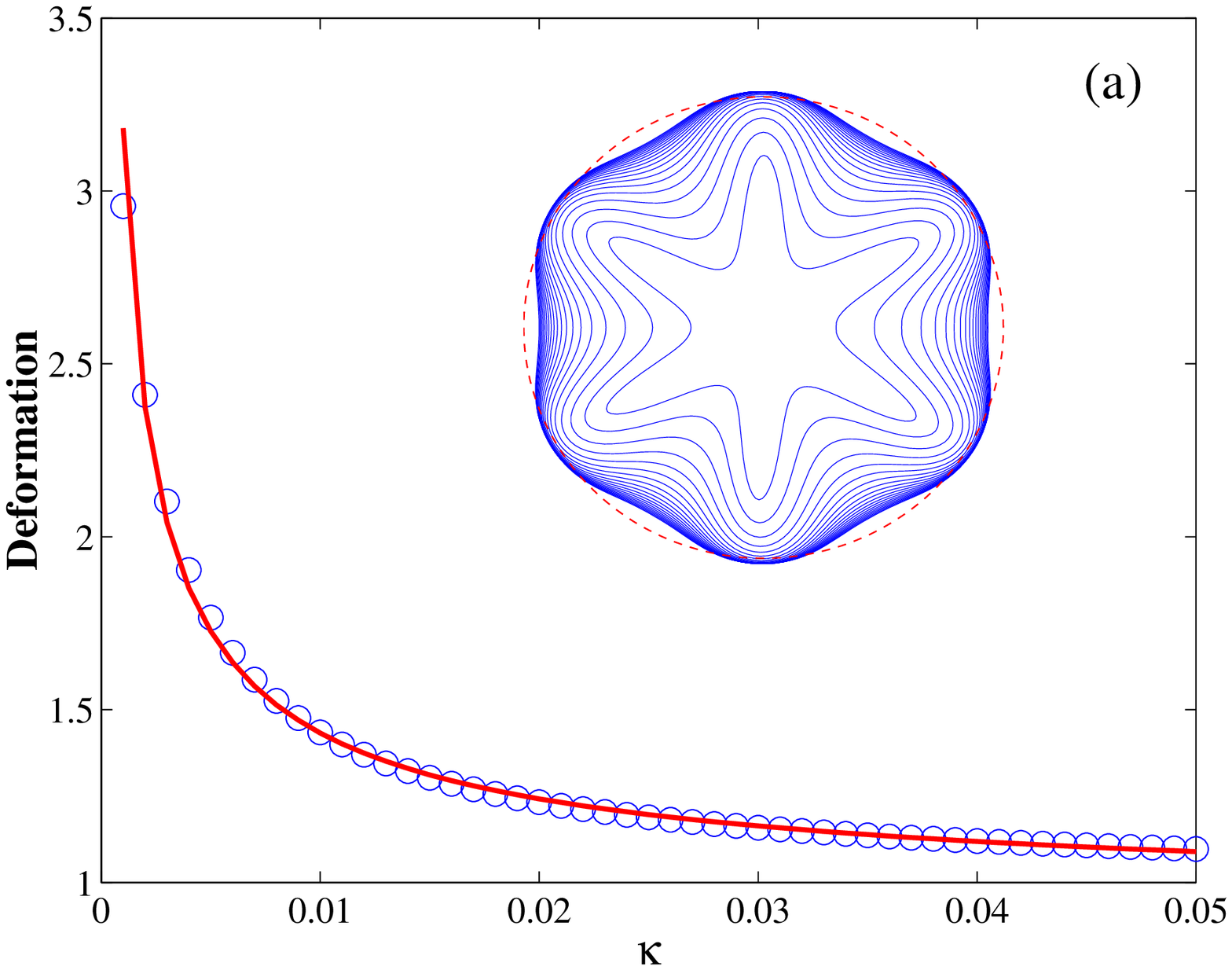, width=2.7in}}
\centerline{
\epsfig{file=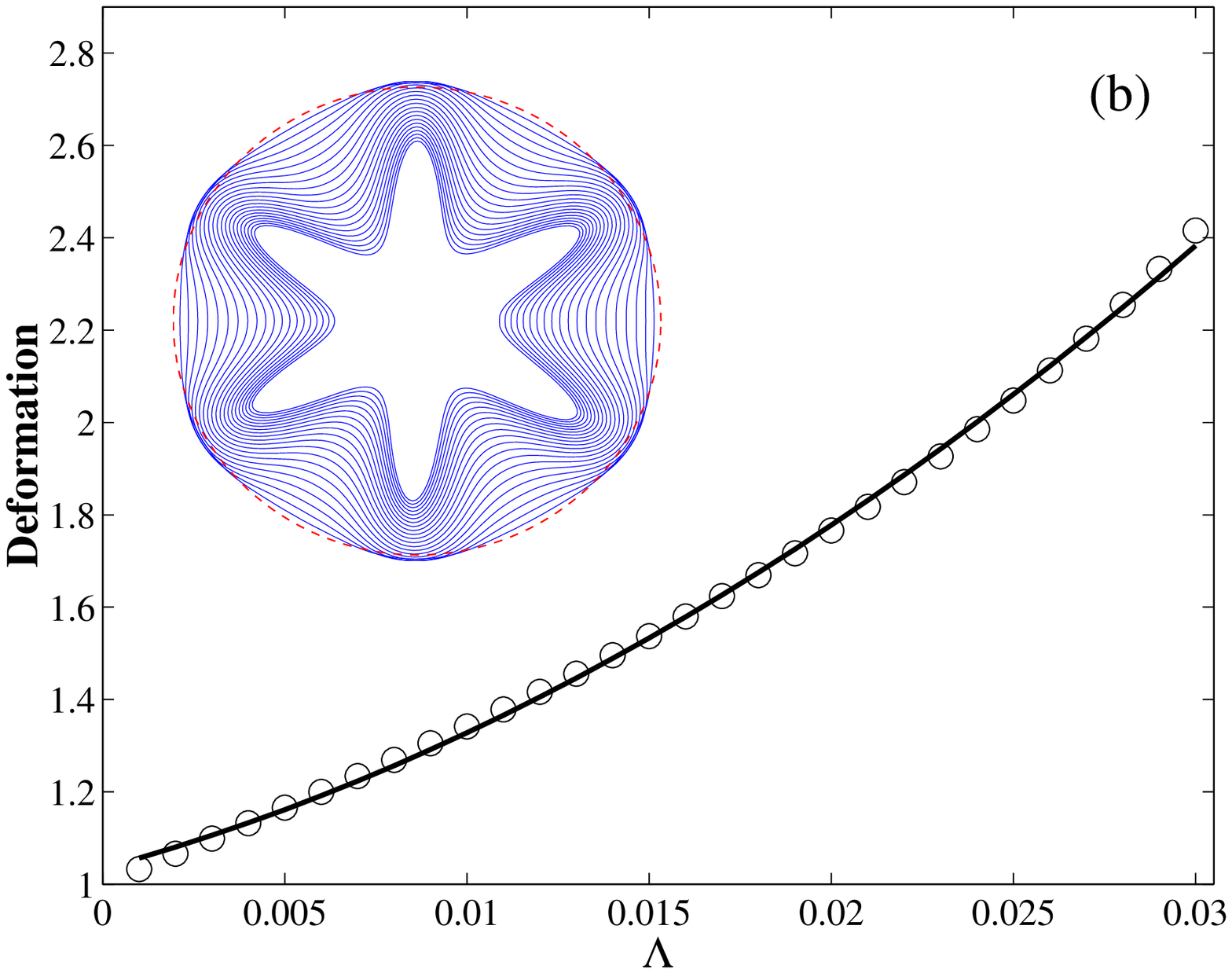, width=2.7in}}
\end{figure}

\begin{figure}
\caption{Phase separation and deformation on spheres and tori. (a) 
Axially symmetric sphere. (b) Azimuthally symmetric sphere. (c) Axially 
symmetric torus. (d) Azimuthally symmetric torus. Shades of gray 
correspond to different magnitudes of the order parameter $\phi$.}
\vskip 1\baselineskip 
\centerline{
\epsfig{file=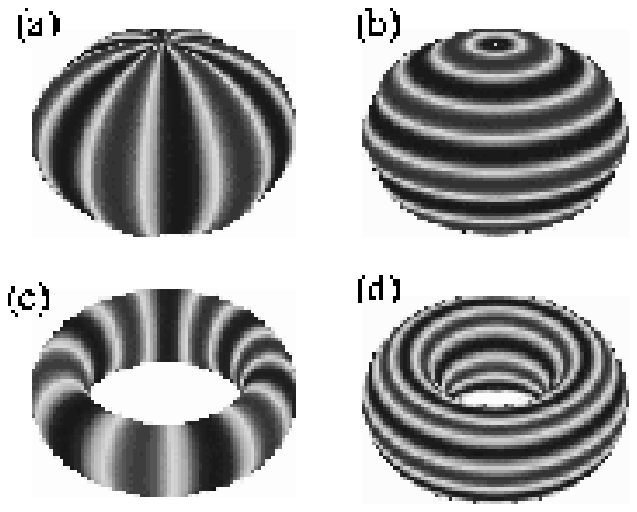, width=2.8in}}
\end{figure}

\end{document}